# When is Software a Medical Device?
- Understanding and Determining the "Intention" and Requirements for Software as a Medical device in EU law

K. Rosager Ludvigsen*

Shishir Nagaraja**

Angela Daly***

*The role of software in society has changed drastically since the start of the 21st century. Software can now partially or fully facilitate anything from diagnosis to treatment of a disease, regardless of whether it is psychological or pathological, with the consequence of software being comparable to any other type of medical equipment, and this makes discovering when software must comply with such rules vital to both manufacturers and regulators. In lieu of the Medical Device Regulation we expand on the idea of intention, and identify the criteria software must fulfil to be considered medical devices within EU-law.*

## I. Introduction

Software and applications play an ever-increasing role in healthcare and wellness. This is evident with the increased use of IoT devices, applications on smartphones for both health professionals and consumers, and even more sophisticated software used by health professionals. This presents a dilemma for lawmakers as to when this software should be regulated in the same way as other medical equipment or 'medical devices' (MD). Software can potentially harm or affect humans in the same way as physical equipment, so there is no doubt that is must be regulated with the same degree of discrepancy.

In the field of MD regulation, the EU and the US are the two biggest players internationally. In this paper, we concentrate on the EU, and attempt to identify when software can be considered a MD. There are a range of reasons why it must be clear when software is and is not a MD, with the most urgent being the increased risks that software used as a MD pose [1]. Software requires security to not be vulnerable to adversaries of any kind, and failure risks the mental and physical health of human beings. Proper security is sadly not the standard[2], and several authors have shown clear issues with security in MD[3]. This is an equally big risk when collecting sensitive data in the form of biometrics because the risk of a breach is that much greater. Security failures[4] can cause further harm when the software is essential for implantable and devices that otherwise affect the physical health of a

---

\* PhD student, Department of Computer and Information Sciences, University of Strathclyde.
\*\* Reader, Department of Computer and Information Sciences, University of Strathclyde.
\*\*\* Reader, Strathclyde Law School, University of Strathclyde.
[1] Kevin Fu et al., "Safety, Security, and Privacy Threats Posed by Accelerating Trends in the Internet of Things," *Computing Community Consortium*, 2020.
[2] Kevin Fu, "Trustworthy Medical Device Software," *Institute of Medicine Workshop on Public Health Effectiveness of the FDA 510 (K)* 510 (2011): 1–20, http://www.cs.ucsb.edu/~sherwood/cs290/papers/fu.pdf.
[3] Pardeep Kumar and Hoon Jae Lee, "Security Issues in Healthcare Applications Using Wireless Medical Sensor Networks: A Survey," *Sensors* 12, no. 1 (2012): 55–91, https://doi.org/10.3390/s120100055; D Iana P T Obón, T Iago H F Alk, and M Artin M Aier, "Context Awareness in WBANS: A Survey on Medical and Non-Medical Applications," *Wireless Communications for E-Health Applications*, no. August (2013): 30–37.
[4] Whether caused by adversaries or non-adversarial failures, like accidents, is irrelevant if damage is caused.

human (such as pacemakers), and these are also known to have poor defences[5]. Preventing malware and other harmful features from being present should be required, but there is no explicit legal or practical requirement for software in general or MD software specifically[6], with no certification or authority investigating or inspecting this specific issue. There are also no requirements for making sure that what software contains is validated and correct[7], and that the software is created with the oversight of a professional[8]. Before considering whether the existing legal requirements take account for these troubles, current research on guidance derived from EU law for software is promising[9], in the sense that they do consider security failures[10]

This is an issue seen during the COVID-19 pandemic and measures to address it includes software. The European Commission issued a Recommendation *'on a common Union toolbox for the use of technology and data to combat and exit from the COVID-19 crisis, in particular concerning mobile applications and the use of anonymised mobility data*'[11]. This was released after the WHO declared COVID-19 to be a pandemic on the 11 March 2020[12]. Many countries have implemented contact tracing of virus infections, using both manual and digital methods such as apps due to the widespread adoption of smartphones and wearables.

Among the many consequences of the virus, was the postponement of Regulation 2017/745 – the Medical Device Regulation (MDR)[13], which is also a primary focus of this paper. It was supposed to have entered into force on 26 May 2020, but is now scheduled to apply from the same date in 2021. Before it was postponed, the European Commission created the aforementioned Recommendation on COVID-19 data. In its preamble 13, it states that the future regulation (MDR) as well as the still-in-force Medical Device Directive (MDD)[14] might apply to some of the mobile applications that could be used for diagnosis, prevention, monitoring, prediction, prognosis, treatment or alleviation of disease during the pandemic. This could include self-diagnosis software, and the Commission directly asks for stakeholders to consider whether this software falls within the scope of the MDD and MDR.

This paper focuses on this ambiguity implicit in the questions the European Commission asks in its Recommendation and which is at the heart of EU MD law - namely the situations in which software falls within the scope of the MDR. We examine how this was addressed by MDD, and how it will be addressed by MDR.

---

[5] Carmen Camara, Pedro Peris-Lopez, and Juan E. Tapiador, "Security and Privacy Issues in Implantable Medical Devices: A Comprehensive Survey," *Journal of Biomedical Informatics* 55 (2015): 272–89, https://doi.org/10.1016/j.jbi.2015.04.007.

[6] Maged N. Kamel Boulos et al., "Mobile Medical and Health Apps: State of the Art, Concerns, Regulatory Control and Certification," *Online Journal of Public Health Informatics* 5, no. 3 (2014): 1–23, https://doi.org/10.5210/ojphi.v5i3.4814.

[7] Ma R. Cantudo-Cuenca et al., "A Better Regulation Is Required in Viral Hepatitis Smartphone Applications," *Farmacia Hospitalaria* 38, no. 2 (2014): 112–17, https://doi.org/10.7399/FH.2014.38.2.1125.

[8] Daniel J. Stevens et al., "Obesity Surgery Smartphone Apps: A Review," *Obesity Surgery* 24, no. 1 (2014): 32–36, https://doi.org/10.1007/s11695-013-1010-3.

[9] Lisa Parker et al., "A Health App Developer's Guide to Law and Policy: A Multi-Sector Policy Analysis," *BMC Medical Informatics and Decision Making* 17, no. 1 (2017): 1–13, https://doi.org/10.1186/s12911-017-0535-0.

[10] The issue with most of the guidance however, is that it has ambiguous legal value, and in practice it is not generally enforced, and as a result it is left up to manufacturers to decide whether to follow it or not (if they are even aware of it in the first place).

[11] Commission Recommendation C(2020) 2296 of 8.4.2020.

[12] https://twitter.com/WHO/status/1237777021742338049

[13] Regulation (EU) 2017/745 of the European Parliament and of the Council of 5 April 2017 on medical devices, amending Directive 2001/83/EC, Regulation (EC) No 178/2002 and Regulation (EC) No 1223/2009 and repealing Council Directives 90/385/EEC and 93/42/EEC.

[14] Council Directive 93/42/EEC of 14 June 1993 concerning medical devices.

The analysis goes through relevant legislation and concepts, including a framework for understanding a core requirement for MD that has not been analysed in detail. Thereafter, we describe how software as MD can be identified, and develop a decision diagram that quantifies it.

## II. Legal sources and software and its use in the medical industry

Use of surgical robots, pacemakers, automated insulin dispensers and a large range of other medical equipment that all make use of software in one way or another, is at an all-time high. In an European context, this is not unfamiliar or unusual, but the Medical Device Regulation is. From late May 2021, all MD in the EU must comply with it. Issues arise due to the fact that it is a regulation. Despite this, the use of software is very distinct when it comes to MD, because of its lack of a physical presence. It may show its existence physically, if is part of a device (an accessory) or if it is necessary for it to function[15], and it may also be standalone and yet a MD.

Increased use warrants scrutinization of its structure and usage. This fits how physical MD are inspected and considered as it stands. There is therefore an angle of security and privacy on top of the MD perspective, as there would be an engineering/pharmaceutical/surgical angle to automatic insulin pumps or surgical robots respectively. Interdisciplinary approaches are therefore not novel in this field. Software as and with MD fill a role that is been chosen[16] to be inseparable from and with the devices. Increased use of IoT or robots[17] also come with increased risks in the form of security and privacy breaches. A security breach may include physical harm on the user, which was discussed extensively in an earlier paper[18]. Likewise, privacy breaches come with their own consequences in the form of theft of personal data, either to sell or use for more sinister purposes. Both of these risks are carried into the MD field from their origin, and this paper is also an attempt to show how this shapes out in a legislative sense.

### 1. Medical Device Directive

MD are as of the time of writing no longer governed by the MDD. By the end of May 2021, it will have become obsolete as a source of law, but it is worth describing because of its similarities to the MDR and its historic significance.

The term software is only mentioned twice in the Directive. One of these places is in art. 1(2), where software is included if it is 'necessary for proper application intended by the manufacturer to be used for human beings'[19]. Any software then has to fit into being an *'instrument, apparatus, appliance, material or other article'*, and be intended for certain purposes. These are:

1. Diagnosis, prevention, monitoring, treatment or alleviation of disease,
2. diagnosis, monitoring, treatment, alleviation of or compensation for an injury or handicap,
3. investigations, replacement or modification of the anatomy or of a physiological process,
4. control of conceptions.

These may not achieve their main goals in/on the human body by pharmacological/immunological/metabolic means[20], and this is because the devices may be regulated by the *in vitro* MD regulation[21]. Outside of fulfilling those requirements, software initially has to be 'intended by the manufacturer to be used on human beings'. We come back to what this means later. Software must therefore be treated and evaluated after the same rules as every other MD, if they are covered by the jurisdiction of MDD. Software can also be considered an accessory to a MD, and these must fulfil the same requirements as them, see art. 1(2)(b). Unlike the regulation, the MDD is a directive, which means that there exists a different implementation of the directive in each EU member state[22]. We recognize that some of the criteria are similar compared to what we see in the regulation, but because of how the wording applies in the future (directly unlike before), they will not be used the same way in practice. There is no clear article on scope and subject, but art. 2 comes close. It shows that a primary purpose of MDD is the placement onto the market of MD, but only if they do not compromise the health of the patients, users or other persons when used for their intended purposes. The rest of the structure resembles that of MDR, in a shorter and more concise form[23]. Since the MDR is more comprehensive, we will go through it in more detail than the Directive[24].

---

[15] And perhaps be an accessory or MD, but this is up for debate.
[16] In the past, many tasks were solved without software, so using them in an increased manner is a choice taken by the manufacturer.
[17] This includes surgical robots, but also devices that can lift individuals out of beds, or anything that can be programmed to move.
[18] K Rosager Ludvigsen and Shishir Nagaraja, "Dissecting Liabilities in Adversarial Surgical Robot Failures : A National (Danish) and European Law Perspective," *Prepublication*, 2020, https://t.co/0o7hVxYpcn?amp=1.
[19] In a rephrased fashion, the same purposes have to be fulfilled in the directive as from art. 2(1) in the MDR.
[20] But they can assist in it, but cannot be the principal way to achieve it.

## 2. Medical Device Regulation

An EU regulation is used directly through its literal wording, unlike directives which have to be implemented into national law[25]. This means that the MDR must be understood as is and per its wording. The scope of the regulation seen in art. 1(1), is to define rules for MD in the three senses of 'placing on the market/making available on the market/put devices into service' for human use and accessories[26]. This means that the subjects and the core focus of the regulation are the devices, and therefore also their manufacturers. They must as subjects to the regulation therefore play a substantial part in complying with the rules, self-evidently identifying and loyally fulfil their duties. The jurisdiction of the MDR is potentially worldwide. If we go back to art. 1(1), any manufacturer that wants to enter the European Single Market must conform with the MDR. The term software is mentioned in the regulation[27] but not treated separately or has any specific articles concerning it. The general rules for MD therefore apply to them. Article 1(1) on subject matter and scope in the regulation, does not literally exclude software as being considered a "medical device for human use", and they are not excluded in art. 1(6) either. In the regulation's definitions in art. 2, software is mentioned as a MD if it used for either or several of these purposes[28]:

1. Diagnosis, prevention, monitoring, prediction, prognosis, treatment or alleviation of disease,
2. diagnosis, monitoring, treatment, alleviation of, or compensation for, an injury or disability,
3. investigation, replacement or modification of the anatomy or of a physiological or pathological process or state,
4. providing information by means of in vitro examination of specimens derived from the human body, including organ, blood and tissue donations.

As with MDD, software can also be an accessory to a MD, see art. 2(1). The MDR makes a sharp division between software intended for use with or as a MD and software for general purposes, see preamble 19. If not intended to, 'generic software', which we define as software for general purposes, can never be an accessory or a MD. But if it is specifically intended by the manufacturer to be used as such (but is not generic), it may be considered a MD or an accessory if it fulfils the other requirements. Software which is MD or is an accessory must also have the CE marking, see art. 20.

---

[21] Regulation (EU) 2017/746 of the European Parliament and of the Council of 5 April 2017 on in vitro diagnostic medical devices and repealing Directive 98/79/EC and Commission Decision 2010/227/EU. The same distinction applies for the MDR.
[22] This is not the intention, as harmonization of the rule sets are preferred, but merely a statement of the obvious, with the opposite of lack of harmonisation being the case.
[23] Only 23 articles compared to the 123 of the MDR.
[24] This paper is not a review or comparison of the two, other than the parts that are relevant to software.
[25] See article art. 288 in the Consolidated Version of the Treaty on the Functioning of the European Union.
[26] This is deeply contrary to art. 2 in the MDD, since MDR does not care about safety or health, at least in its scope article.
[27] Preamble 19, art. 2(1)(4)(25)(26) and in the annexes.
[28] The resemble those seen in the MDD, but are slightly different in wording.

In the view of the public and manufacturers, the character of the MD is to be assessed before it is released. This is done with classes that designated increased risk. MD are divided into class I, IIa, IIb and III, see art. 51(1). The determinations of these classes are defined in Annex VIII in the MDR. The difference in class can be seen as the danger it poses to those they are used on (be it patients or users), and to the special rules that are given to the higher classes. For class III, this would include art. 27/32/52/54/55/61/86/105[29]. Software is mentioned in section 3.3 of Annex VIII, which dictates that if it drives or influences a device it shall have the same class. This is the main rule. But if the software is independent, it has to be classified in its own right. Specific rules follow these two principles. Rule 11 in Annex VIII assumes at first, that all software that is also a MD is class I[30]. The exception to this is software which is used to take decisions on diagnosis or has therapeutic purposes. These are instead considered class IIa. And the exception to the exception would be if the software is capable of causing serious deterioration of the health of a person, which makes it class IIb, or if it is capable of causing death or irreversible deterioration to the health of a person, in which case it is class III. If the software monitors physiological processes, it is class IIa, but if it can cause immediate harm to the patient during its use, it is to be considered class IIb.

MD in the EU are enforced by nationally appointed regulators, see art. 101. The Medical Device Coordination Group (MDCG) as seen in art. 103, is the organ that facilitates cooperation between the different regulators, but it is not by itself the central authority. The authorities can make use of a wide array of enforcement tools, including forceful withdrawal and banning of the MD, see art. 10(14). Their practice is not public and/or does not exist, which makes prediction of regulatory behaviour difficult. EU regulators do not always assess the conformity of the devices initially. This is at first left to something called a 'notified body', see art. 35 and 36. These are usually private companies, which are given the competence to access MD[31].

## 3. Meddev 2.1/6 and MDCG 2019-11[32]

Issued with the MDD, is the 'Guidelines on the Qualification and Classification of Stand Alone Software used in Healthcare within the Regulatory Framework of Medical Devices', called Meddev 2.1/6 for short. We will only bring up points which are important to our discussion, as the guidance also contains several outdated passages.

Issued as a guidance before MDR applied, 'Guidance on Qualification and Classification of Software in Regulation (EU) 2017/745 – MDR and Regulation (EU) 2017/746 – IVDR', called MDCG 2019-11, is the updated guidance on software as MD. We will bring out parts that are relevant for analysis from this as well.

Because Meddev 2.1/6 was built on MDD, it did not have many specific articles to draw inspiration from. It does add that generic software should not be considered a MD, if it is part of several software modules in one unit[33]. Every evaluation of such software must be individual (for each type of software), and so that one module which is a MD does not make the entire system one as such. A central limitation of this guidance is that it only applies to standalone software[34] and not software incorporated in MD such as those seen in surgical robots. But this does not limit its role for inspiration for interpreting the MDR in the future, together with the new guidance. The guidance assures that the intent of the manufacturer plays a central role, regardless of the name of the software[35]. This can be interpreted as the situation where the software may be called something in regards to healthcare, but its intent from the manufacturer was purely to act as something informal. The choice of operating system that the software runs on does not affect the evaluation of the software either, and the risk of malfunction of software run in a medical environment does not make into a MD because of it[36].

---

[29] Some of these articles contain other rules, like art. 32, but they are listed because they include special obligations for manufacturers of MD that are class III.
[30] Such as Google's latest attempt to launch a Disease Diagnosis Software in the EU, which is as of the time of writing Class I, see https://blog.google/technology/health/ai-dermatology-preview-io-2021/amp/.
[31] While this is intriguing, it does not play a role in this analysis and will not impact this paper.
[32] Guidance has no enforcement article in the MDD or MDR. This means it is self-regulation, but it may play a role in the practice of national authorities.
[33] See p. 20 – 21 of Meddev 2.1/6.

The guidance includes a "decision diagram to assist qualification of software as medical device"[37]. The MDCG 2019-11 has a slightly updated diagram[38] that we will also use as inspiration. We choose to make use of this decision diagram for our own proposed set of requirements later, but first we need to see how each point relates to the rest. It has to be noted that even if the diagram states, that if the software does not fulfil the criteria, it is not covered by the MD directives[39], we interpret that as meaning that the software is not considered a medical device. This has been fixed in the new guidance.

Comments on the central parts of the diagram from Meddev 2.1/6 are the following:

1. First point relates to whether the software can be defined as software in the guidance. Related to the MDR, there is no stringent definition of this. Clearly, only software can be considered software.

2. Second point relates to whether the software is standalone or not. If it is not, it can either be part of a MD or not covered by the MDD.

3. Third part defines, that if the software merely acts on data from storage, archives, communication or simple search, it is not a MD.

4. Fourth part postulates, that if the actions of the software are not for the benefit of individual patients, it is not a MD. Note that neither MDD nor MDR has the benefit of patients as their main purposes[40].

5. Fifth part says, that if the software does not fit a purpose put forth in art. 1(2)(a) of the MDD, it is not a MD, and it is if it does. But even if it is not a MD, the software can still be considered an accessory, which would lead to the same requirements as that of a MD.

6. Sixth part says, that to be an accessory, the software must fulfil art. 1(2)(b), which means that it must be intended specifically to be used with a MD by the manufacturer. Otherwise, it is not a MD or an accessory at all.

As the keen reader would have noticed, that while the whole diagram is interesting for inspirational and historical reasons the fourth point stands out. The MDR does not work with whether it benefits individuals, it is not even mentioned literally in the text. It instead considers whether the devices are for human use. MDCG 2019-11 repeats several of these requirements in its diagram and text, but it is notable different.

---

[34] This term is no longer used, see MDCG 2019-11, footnote 2.
[35] P. 8 of Meddev 2.1/6.
[36] Supra, P. 9.
[37] Supra, P. 10.
[38] P. 9 of the MDCG 2019-11.
[39] There exists a distinction between the MDD and directive 98/79 on in vitro medical devices, and vice versa with the MDR and its in vitro sibling.
[40] But MDD does it have it in art. 2 somewhat as we showed.

This guidance does not view software in modules, and instead focuses on if it fulfils the requirements to be a MD by itself, if it drives or influences a MD, regardless of where it is, and if used explicitly for a MD related purpose by staff/lay persons[41][42]. The guidance further simplifies the decision diagram from Meddev 2.1/6[43], and differs with this and the lack of reference to the MDR's art. 2(1) – instead opting to refer to its own definition of Medical Device Software[44].

### 4. Case law, C-329/16[45][46].

Before we create our framework, we need to analyse one central case for the use of this kind of guidance and software as MD in general. A fair amount of cases has been decided on by the ECJ in regards to MD. However, most of these do not concern themselves with software, but we do have one that has influence on both MDD and MDR. The case is what is a preliminary ruling, which means that a national court asks the ECJ about the interpretation of EU law in some aspect. The question asked relates to the Medical Device Directive, not the MDR, but because of how recent the verdict was given (2017), it likely applies directly to the Regulation. The case concerned one question with two core points.

*First*, whether standalone software which gave "medico-social establishment support for determining a drug prescription" could show that it was a MD[47].

*Second,* which asked whether a MD has to act in or on the human body to be considered one[48].

The ECJ answered both in full. The first rests on what the manufacturer intended for the software to be used as, like we see in preamble 19[49] in the MDR, and therefore whether it fulfils one or several conditions in the MDD art. 1(2). If the manufacturer intended for it to be used in this manner, it is a MD. Like now, generic software that is used in medical setting is also clearly not a MD. Since the software supports the doctor with his decision making in a manner that could affect the patient, it must be considered a MD.

The second question is answered in reference to itself and the now preamble 19, in that there is no difference whether the software as a MD has contact with a human or not, but rather what the intention of the software is.

The case may seem obvious with event of the MDR, but because the MDD was a directive, it relied on national implementation and interpretation, and while this case came through relatively late, it cemented the importance of the manufacturer's intention in the partially fragmented state that the MDD was in.

---

[41] Example is insulin injection, whether via electrical pump or manually via syringe.
[42] P.7 of MDCG 2019-11.
[43] Supra, P. 9.
[44] Supra, P. 7.
[45] Case C-329/16 SNITEM, Judgment of the Court (Fourth Chamber) of 7 December 2017.
[46] A brilliant analysis of this verdict can be read in - Timo Minssen, Marc Mimler, and Vivian Mak, "When Does Stand-Alone Software Quality as a Medical Device in the European Union? - The CJEU's Decision in SNITEM and What It Implies for the Next Generation of Medical Devices," *Medical Law Review* 0, no. 0 (2020): 1–10, https://doi.org/10.1093/medlaw/fwaa012.
[47] C-329/16, para 20.
[48] Supra, para 20.
[49] The equivalent is preamble 6 in directive 2007/4, which amends the MDD.

## III. The Intention of the Manufacturer

In this section we create a framework to expand the idea behind manufacturer's intention for their software to be MD. This is necessary because of the intrinsic values that software defines itself from, compared to other types of MD. A scalpel or a syringe is not capable of collecting personal data or independently control itself, but many types of software are.

### 1. Preliminary Comments

The intention is central as to whether the MDR applies in the first place, which is seen in the definition of MD in art. 2(1), and requires the manufacturer to have intended for their product to fulfil one of these purposes. This was further emphasized in C-329/16 as we just showed. In MDR this is further defined in art. 2(12), which states that the labelling, instructions for use, promotional material or statements made by the manufacturer in clinical evaluation is considered "intended use"[50].

But how we are supposed to deduce intention, besides from when it is written literally, is not clear from the MDR, nor its preparatory materials nor past practice or the guidance. Art. 2(12) does not go into actual use or how the manufacturers designed the software.

We therefore propose a framework which can aid with this determination and increase clarity and certainty for software manufacturers and ensure consistency for regulators.

The reason we assume that degrees or other abstractions of the intention is needed, is because of the special attributes of software. Unlike other medical equipment, the use and the design of software can be deceptive. This refers to the hidden layers and the ethereal nature of software. Physical appliances have no hidden features like data collection or risks of cybersecurity breaches, but software does, so this framework is made to aide in identifying these as well.

For the MDR to be effective, it must cover all MD possible, and this framework may help all stakeholders realise that. Of course, this framework lists what we consider to be apparent indicative sources to determine the manufacturer's intention in regards to software as MD, and can therefore not be considered an exhaustive list.

The other reason why we must include more considerations into intention, is because it serves as the easiest way to prevent the MDR from applying to one's software. If we go by the definition in art. 2(12), software that does directly say that it is used for purpose from art. 2(1), will not be considered a MD. But the software may very well be used by laymen or medical professionals[51] for treatment, or it may be an app of the Google Play store that would fulfil art. 2(1) had the manufacturer labelled it as a MD. It is this grey area we want to shine a light on and quantify.

### 2. Framework for manufacturer's intention to qualify software as a medical device

Firstly, we must assume that the MDR did not intend for the manufacturer to be able to circumvent the entire regulation by merely stating that their software is not a MD. This should instead be decided by the purposes laid out in art. 2(1) as is the intention. If the regulation allowed easy circumvention, it would not be able to fulfil its goals in art. 1(1), and be rendered positively legally redundant[52].

---

[50] Refers to tests conducted by notified bodies, not the national authorities initially, see art. 61 in the MDR.
[51] MDCG 2019-11 does predict and see these types of software as being intended to be a MD, but does not specify how, when and where, se p. 7.
[52] We are aware that the MDR (and MDD) are part of the overarching product rules in the EU, but this does not hinder special rules or considerations for highly specialised products as such.

Secondly, we must expand the term "intention" to the *direct intention* seen in publicly available documentation and marketing materials issued by the manufacturer that concern the software[53]. But it must also include the manufacturer's *indirect intention*, which is seen in what the software is capable of, what data it retrieves or measures, and what kind of analysis or lack thereof the software is able to do. The indirect intention can therefore be considered as to what the software is capable of and what it does in practice. This follows the idea behind an accessory - software specifically made to support/enable other MD to function are accessories and therefore covered by the same MDR rules. Accessories are defined by their abilities, not just what they are said to be able to do, and using this terminology on MD as such is worth considering.

Indirect intention has to viewed as the actual capabilities of the software, regardless of what is stated in publicly available sources.

We consider that in the case of conflict between the two types of intention, if either indicates the intention of the manufacturer for the software to be a MD, the affirming intention will prevail. This applies regardless of whether it is direct or indirect, as software that is claimed to be a MD (direct intention) can still at fail at being it in practice (indirect intention) and vice versa, but it would still be considered a MD for the purposes of the MDR.

It may also not be a MD because it does not fulfil any of the purposes in art. 2(1), even if the manufacturer intended for it to be so.

### 3. The Framework
Firstly, we propose the sources for discerning direct intention:

1. Information from marketing materials. If the manufacturer states that the software is a medical device, or claims it fulfils one of the purposes in art. 2(1), direct intention can be established.

2. Information from internal documentation. If the manufacturer states that the software is supposed to fulfil one of the purposes in art. 2(1) in its internal documentation to which the European regulators or other public authorities have access due to art. 10 or other legal provisions (e.g. in national public law) direct intention can be established.

3. Informal information sources. If manufacturers' representatives have said elsewhere, or if it stated in search systems through such mediums as tags that the software fulfils one of the purposes in art. 2(1), direct intention can also be established.

Secondarily, we propose these as sources for discerning manufacturers indirect intention:

1. Data gathering practices. If the software gathers or measures data that is relevant for fulfilling the purposes in art. 2(1), an indirect intention can be established. This can be biometrical data about, as well as health records of, an individual.

2. Software specifications. If the software is designed and functions as if it was a medical device, or if it only operates in a setting where it would be considered an accessory to a medical device, an indirect intention can be established.

3. Data analysis. If the software, as part of its purpose, requires personal data to be analysed to reach results that resemble the purposes in art. 2(1), an indirect intention can be established.

---

[53] This is the idea laid out by MDCG 2019-11 as we showed earlier, and art. 2(12) to some extent, as well as one novel angle.

An illustration of the Framework can be found in the Annex, figure 1.

# IV. Requirements for Software to be a Medical Device

In this section we make a set of more direct requirements for when software is considered a MD or not in the regime of the MDR. We make the requirements fit the MDR, because it is due to be the active, while the MDD is soon to be inactive.

We also explain how software should be classified in this context as well.

## 1. Overarching concepts

The first three distinctions that are necessary, is to make the intention of the manufacturer clear, ask whether the software is generic and whether the software is an accessory.

1. If the software is not intended for one or more of the medical purposes in art. 2(1), it cannot be considered a MD. This is subjective, and we showed in our framework that the intention can be direct or indirect. Because of the wording in the MDR, resolving whether there is intention or not must be done before anything else.

2. If the software is generic and not modified by the manufacturer to be used standalone or as in a module/system, it is not a MD[54].

3. If the software is intended by the manufacturer to be an accessory to a MD, either directly or to assist it, the software cannot by itself be considered a MD. It is therefore considered an accessory instead.

We know from C-329/16, that there exists a fourth requirement.

4. If the software is not used for/on human beings, it cannot be a MD.

## 2. Considerations for the requirements

The decision diagram used for standalone software in Meddev 2.1/6, fourth point, distinguishes between software that stores, archives, communicates or searches in data and software that does other actions. There is no basis for this in the regulation, but from the points above the idea is supported, since purely storing/archiving/communicating and searching does not fulfil any of the purposes, even if the manufacturer may intend it to do so.

As these are interpretation points, certain considerations have to be taken for each. All points refer to software that specifically is intended to be used for such purposes, this rules out software that merely facilitates and is specifically not intended for it. But it also catches software that merely monitors disease/disability[55].

Alleviation of disability does not exist, but compensation is possible - which means that software that for example is used to control an artificial limb is either a MD or an accessory to a MD. It even includes replacement of limbs or organs as part of "replacement or modification", even if the software merely calculates movements or facilitates pumps. This part is extremely wide due to it covering all physiological/pathological processes or states.

When "used" is written, we think back to C-329/16, that in line 30 and 32, assures us that the software does not need to physically be "used" on the human body. This again broadens the range of software that may then be considered MD. But as broad as it sounds, we still want to include a criteria that covers and forces the software to be used on/for humans, even if there does not exist a physical requirement as such, which is in line with C-329/16.

---

[54] We have defined this negatively in the diagram, see Figure 1 in the Annex.
[55] See art. 2(1), 'monitor' mentioned in the two first sets of criteria.

The framework of intention plays a big role as to whether we can even start using the requirements. While the MDR does not explicitly write that the regulators are to search and survey the software market to discover whether software could be a MD or not, they are not prevented in doing so. Their search could therefore include all software clearly marketed as such, and all software that clearly acts as a MD. This kind of search would be part of the obligations of regulators[56]. We illustrate the requirements discussed above as a decision tree, which is figure 2 in the Annex.

## VI. Concluding remarks

Software is here to stay in MD regulation, but the event of horizon that is apps and IoT is arriving fast. With our suggestion to expand the idea and terminology so that intention to includes what software does in practice, we hope that the MDR can cover and increase the quality of software as MD in the future, both in the EU and the world as a whole.

## VII. Next Steps

This article is limited in what it actually can study, which is why papers that explore what is outside of our scope are needed. This could include analysis of specific genres of software, and how they would be treated by the MDR. Specifically, applications that focus on physiological diseases or disabilities or purely accessories, to show how the requirements framework is applied in practice. It could also be practice papers that analyse one or several regulators on a European level, with interviews/unique insight or the like, which are very needed and wanted. Furthermore, policy papers that address the enforcement issues we have brought forth would be usable as well, as would some that could show a different interpretative angle towards the MDR in general. And finally, empirical research on which kinds of software could be considered a MD if one uses indirect intention to determine, through methods such as scraping the Google Play store.

## VIII. Competing interest declaration

There are none.

---

[56] See art. 93 on Market Surveillance for one example.

## Annex

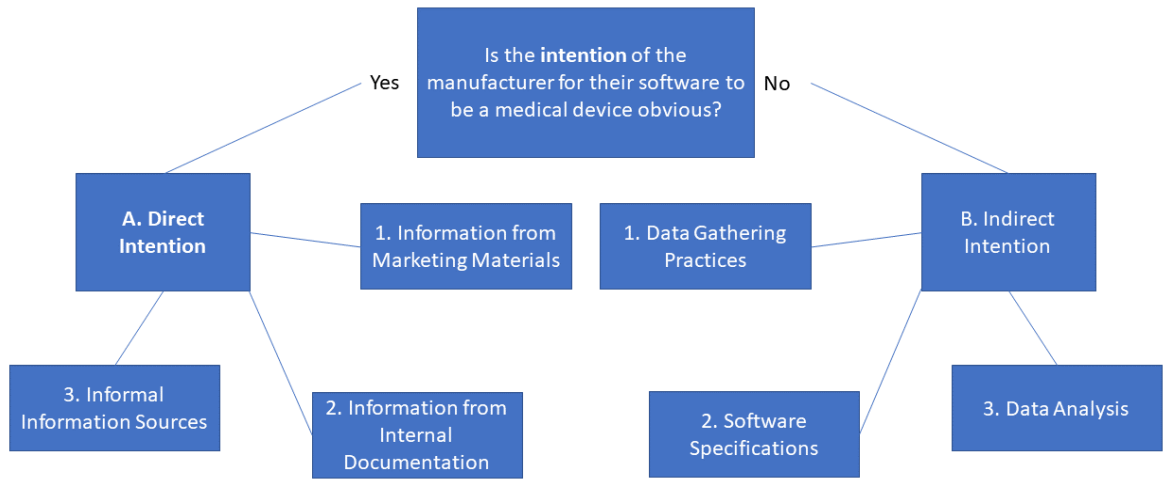

Figure 1 – diagram of the framework of intention.

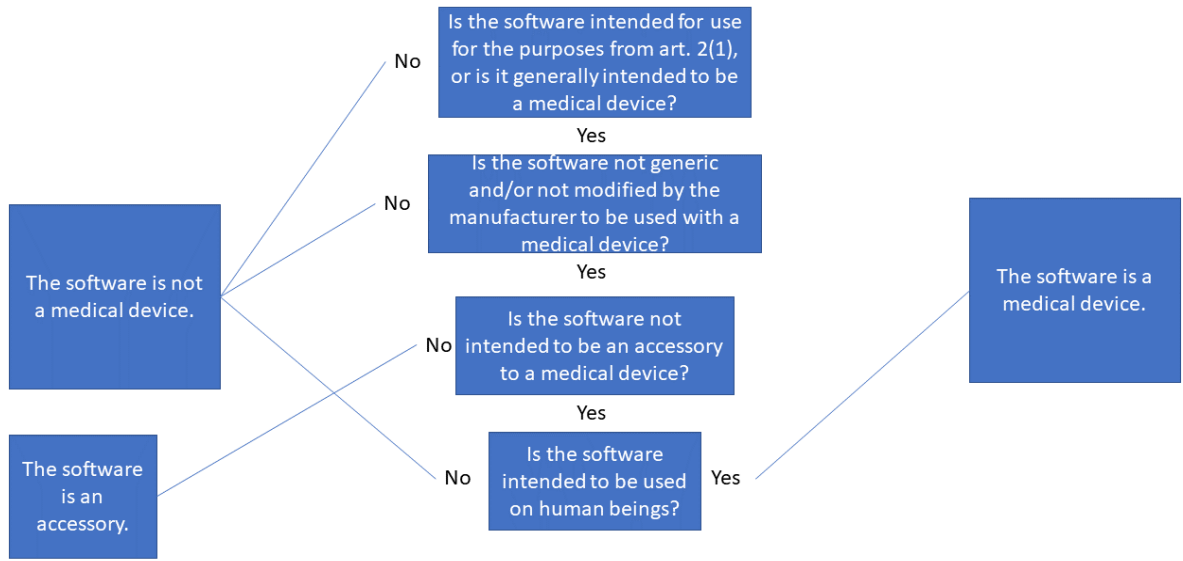

Figure 2 – illustration of the requirements for software to be considered a MD.